\documentclass[
    ,final
  ]
  {aipproc}

\layoutstyle{8x11single}

\begin{document}

\title{Delay time distribution of type Ia supernovae: theory vs. observation}

\classification{97.20.Rp, 97.60.Bw, 97.80.Fk, 98.52.Eh}
\keywords{supernovae, close binaries, white dwarfs, elliptical galaxies}

\author{Nicki Mennekens}{
  address={Astrophysical Institute, Vrije Universiteit Brussel, Pleinlaan 2, 1050 Brussels, Belgium}
}

\author{Dany Vanbeveren}{
  address={Astrophysical Institute, Vrije Universiteit Brussel, Pleinlaan 2, 1050 Brussels, Belgium}
  ,altaddress={Groep T - Leuven Engineering College, K.U.Leuven Association, Andreas Vesaliusstraat 13, 3000 Leuven, Belgium}
}

\author{Jean-Pierre De Greve}{
  address={Astrophysical Institute, Vrije Universiteit Brussel, Pleinlaan 2, 1050 Brussels, Belgium}
}

\author{Erwin De Donder}{
  address={Astrophysical Institute, Vrije Universiteit Brussel, Pleinlaan 2, 1050 Brussels, Belgium}
  ,altaddress={Belgian Institute for Space Aeronomy (BIRA-IASB), Ringlaan 3, 1180 Brussels, Belgium}
}

\begin{abstract}
Two formation scenarios are investigated for type Ia supernovae in elliptical galaxies: the single degenerate scenario (a white dwarf reaching the Chandrasekhar limit through accretion of matter transferred from its companion star in a binary) and the double degenerate scenario (the inspiraling and merging of two white dwarfs in a binary as a result of the emission of gravitational wave radiation). A population number synthesis code is used, which includes the latest physical results in binary evolution and allows to differentiate between certain physical scenarios (such as the description of common envelope evolution) and evolutionary parameters (such as the mass transfer efficiency during Roche lobe overflow). The thus obtained theoretical distributions of type Ia supernova delay times are compared to those that are observed, both in morphological shape and absolute number of events. The critical influence of certain parameters on these distributions is used to constrain their values. The single degenerate scenario alone is found to be unable in reproducing the morphological shape of the observational delay time distribution, while use of the double degenerate one (or a combination of both) does result in fair agreement. Most double degenerate type Ia supernovae are formed through a normal, quasi-conservative Roche lobe overflow followed by a common envelope phase, not through two successive common envelope phases as is often assumed. This may cast doubt on the determination of delay times by using analytical formalisms, as is sometimes done in other studies. The theoretical absolute number of events in old elliptical galaxies lies a factor of at least three below the rates that are observed. While this may simply be the result of observational uncertainties, a better treatment of the effects of rotation on stellar structure could mitigate the discrepancy.
\end{abstract}

\maketitle

\section{Introduction}

Type Ia supernovae (SNe Ia), which are among the most powerful explosions observed in the universe, are events that can occur only in multiple star systems. They are not only critical to the chemical evolution of galaxies (without them, we would for example be unable to explain the amount of iron observed in the solar neighborhood), but are also increasingly being used as distance indicators, or standard candles, in cosmology. Despite this, their origin remains unknown. It is agreed upon that SNe Ia originate from the thermonuclear disruption of a white dwarf (WD) in a binary star, which attains a critical mass close to the Chandrasekhar limit of $1.4$~M$_{\odot}$ (see e.g. Livio 2001). However, the exact formation process, and even the type of systems in which such is possible, is a matter of debate. The two most popular formation channels are the single degenerate (a WD steadily accreting hydrogen-rich material from a late main sequence (MS) or red giant (RG) companion, see e.g. Nomoto 1982) and double degenerate (a super-Chandrasekhar merger of two WDs due to gravitational wave radiation (GWR) spiral-in, see e.g. Webbink 1984) scenario.
\\
\\
To address the question of which of these scenarios is dominant in nature (or both), one can turn to the observational delay time distribution (DTD) of SNe Ia, which is the number of such events per unit time as a function of time elapsed since starburst. Totani et al. (2008) obtain a DTD by observing the SN Ia rate in elliptical galaxies which are passively evolving, and thus equivalent to starburst galaxies for this purpose, at similar (near solar) metallicity but different redshifts. The thus obtained distribution is extended with the SN Ia rate for local elliptical galaxies observed by Mannucci et al. (2005). The result is a DTD decreasing inversely proportional to time and in units (SNuK, SNe per K-band luminosity) which need to be converted (into SNuM, SNe per total initial galaxy mass) in order to be compared to any theoretical model. This conversion factor is obtained from spectral energy distribution templates, but may be subject to uncertainties. Using the observational DTD, it is possible to constrain theoretical models for SN Ia formation in starburst galaxies. For the reason just mentioned, comparisons between theoretical and observational DTDs will mainly focus on the shape of the distributions, and not so much on the absolute values.

\section{Assumptions}

Previous studies have been done on this topic by other groups (see e.g. Ruiter et al. 2009; Hachisu et al. 2008; Han \& Podsiadlowski 2004; Yungelson \& Livio 2000), but the present one specifically focuses on the influence of mass transfer efficiency during Roche lobe overflow (RLOF) in close binaries. This is done with an updated version of the population number synthesis code by De Donder \& Vanbeveren (2004), which computes detailed binary evolution models, without the use of analytical formalisms. Single degenerate (SD) progenitors are assumed to be as given by Hachisu et al. (2008), where regions in the companion mass--orbital period parameter space are denoted, one for the WD+MS channel and one for the WD+RG channel. Systems entering one of these regions will encounter a mass transfer phase towards the WD that is calm enough in order not to result in nova-like flashes on the surface of the WD that burn away any accreted hydrogen, but also sufficiently fast to let the WD reach the Chandrasekhar limit and result in a SN Ia before the companion ends its life. This scenario includes the mass stripping effect, which allows accretors to blow away some of the mass coming towards them, letting some systems which would otherwise have merged escape a common envelope (CE) phase and result in a SN Ia. For the double degenerate (DD) scenario, it is assumed that every WD merger exceeding $1.4$~M$_{\odot}$ will result in a SN Ia.
\\
\\
Certain parameters need to be scrutinized, first and foremost the fraction $\beta$ of RLOF-material which is accepted by the accretor. If $\beta < 1$, mass will be lost from the system and angular momentum loss must be taken into account. This is done under the assumption that matter leaves the system with the specific angular momentum of the second Lagrangian point, since mass loss is considered to take place through a circumbinary disk. Other groups make different assumptions, which can have serious implications for the eventual evolution outcome. Finally, a formalism must be adopted for the treatment of energy conversion during CE phases. For the standard model, the $\alpha$-formalism by Webbink (1984) will be adopted, while another possibility will be considered later on.

\section{Double degenerate evolution channels}

\begin{figure*}
   \centering
   \includegraphics[width=10.75cm]{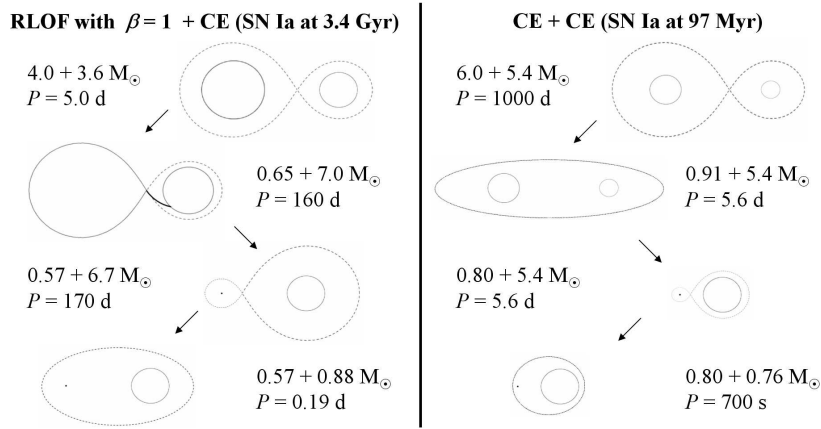}
   \caption{Graphical representation (not to scale) of the two channels typically leading to DD SNe Ia in our population code. Left panel: (conservative) RLOF phase followed by CE phase. Right panel: two successive CE phases.}
\end{figure*}

In our code, there are two evolution channels that can lead to a DD SN Ia, which are represented graphically by typical examples in Fig.~1. In the first channel, the explosion follows an evolution which entails one stable RLOF phase (which is assumed to be conservative: $\beta = 1$), followed by a CE phase. The latter is due to the extreme mass ratio at the start of the second mass transfer phase, and the fact that the accreting object is a WD. In this channel, the resulting system is a double WD binary with a mass of the order of $1$~M$_{\odot}$ each, and with an orbital period $P$ of a few hours. Such a system then typically needs a GWR spiral-in lasting several Gyr, resulting in a SN Ia after such a long delay time. Importantly, if in this channel the RLOF phase is assumed to be totally non-conservative ($\beta = 0$), the system will merge already during that first mass transfer phase, and there will thus be no SN Ia.
\\
\\
The second channel consists of an evolution made up of two successive CE phases. The nature of the first mass transfer phase is a result of the system having an initial orbital period typically two orders of magnitude larger than in the other channel. This makes that the donor's outer layers are deeply convective by the time mass transfer starts, which causes this process to be dynamically unstable. Eventually, after the second CE phase, a double WD binary of about the same component masses as in the first channel is obtained, but with an orbital period of only a few hundred seconds. Such systems require GWR during only a few tens of thousands of years in order to merge, with the SN Ia thus having a total delay time of just a few hundred Myr.

\section{Results and discussion}

\begin{figure*}
   \centering
   \includegraphics[width=10.75cm]{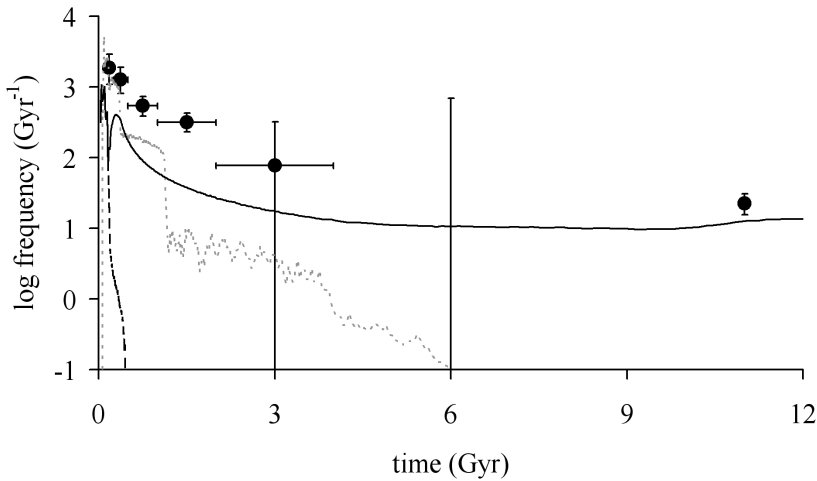}
   \caption{DTDs for $\beta=1$ of the DD (solid black) and SD (dotted gray) scenario, as well as for $\beta=0$ of the DD (dashed black) scenario, using the $\alpha$-formalism for CE. Observational data points of Totani et al. (2008) and Mannucci et al. (2005) (black circles).}
\end{figure*}

The results for the DTD, obtained with the population synthesis code, are shown in Fig.~2. It is obvious that the SD DTD for conservative RLOF ($\beta = 1$) is decreasing much too fast and too soon in order to keep matching the morphological shape of the observational data points after a few Gyr. The SD DTD for totally non-conservative RLOF ($\beta = 0$) hardly deviates from the conservative one shown. The SD scenario by itself is thus incompatible with the observations. We also find most SD events to occur through the WD+MS channel, as opposed to the WD+RG channel. The DD DTD for conservative RLOF matches the observational points in shape, but results in a too low absolute number of events to match them. This may be partially caused by uncertainties in the conversion between SNuK and SNuM, but may also have a physical explanation which will be addressed below. Importantly, most DD SNe Ia are created through a quasi-conservative RLOF followed by a CE phase, not through two successive CE phases. This is also visible from Fig.~2, by comparing the DD DTDs for totally conservative and non-conservative RLOF. In the latter case, the DTD drops dramatically after a few hundred Myr, leaving no DD SNe Ia with a sizeable delay time. This means that the first peak in the DD DTD, present in both cases of $\beta$, contains the events created through a double CE phase, and the second one (the absolute majority of events, but only present in the case of conservative RLOF) those created by a RLOF phase followed by a CE phase. Apart from confirming the aforementioned typical timescales for both channels, and the inability of the channel containing a RLOF phase to produce any SNe Ia if this phase is non-conservative, this also means that in reality a quasi-conservative RLOF (more specifically, $\beta \ge 0.9$) is required to obtain a match in morphological shape between model and observation. This has negative implications for the use of analytical formalisms for the determination of delay times, since such studies typically assume that the lifetime of the secondary star is unaffected by the mass transfer process, which is obviously not true in the case of conservative RLOF.
\\
\\
The next step is a study of the influence of the description of CE evolution. So far, the $\alpha$-formalism by Webbink (1984) was used, which is based on a balance of energy following from a conservation of angular momentum. An alternative is the $\gamma$-formalism by Nelemans \& Tout (2005), which starts from a conservation of energy to arrive at a balance of angular momentum, and which is said to be better for the treatment of systems which will result in a WD binary. The result obtained with this formalism is shown for both the SD and DD scenario and with $\beta = 1$ in Fig.~3. The SD DTD using the $\gamma$-formalism still deviates strongly from the observations, both in shape and number. While the shape of the DD DTD is in agreement with that of the observational data points, it has dropped in absolute number by another order of magnitude as compared to the $\alpha$-formalism. While it is thus not possible to reject the use of the $\gamma$-formalism based on a shape comparison, it seems unlikely that such a large SN Ia rate discrepancy can be explained.
\\
\\
Finally, some considerations are made on the absolute number of SNe Ia. As mentioned before, all considered theoretical models underestimate the observed absolute rate by a factor of at least three at the 11 Gyr point. This might be partially due to the SNuK-SNuM conversion, but a more plausible solution is stellar rotation. If it is so that stars in binaries are typically born with a higher rotational velocity than single stars, for which there seem to be indications (see e.g. Habets \& Zwaan 1989), then it seems likely that a lot of binary components will rotate faster than synchronously on the ZAMS. In that case, they will also have heavier MS convective cores than expected (see e.g. Decressin et al. 2009), which will eventually lead to heavier remnant masses. One will thus obtain heavier WDs, and hence more systems of merging WDs which attain the required $1.4$~M$_{\odot}$ for a DD SN Ia. Figure~4 shows a theoretical DTD obtained for $\beta = 1$ with a 10\% increase in MS convective core mass, and for the SD and DD scenario combined, since there is no reason why both scenarios could not be working together. This DTD agrees well, now both in morphological shape and in absolute number, with the observational DTD by Totani et al. (2008) and with the more recent one by Maoz et al. (2010).

\begin{figure*}
   \centering
   \includegraphics[width=10.75cm]{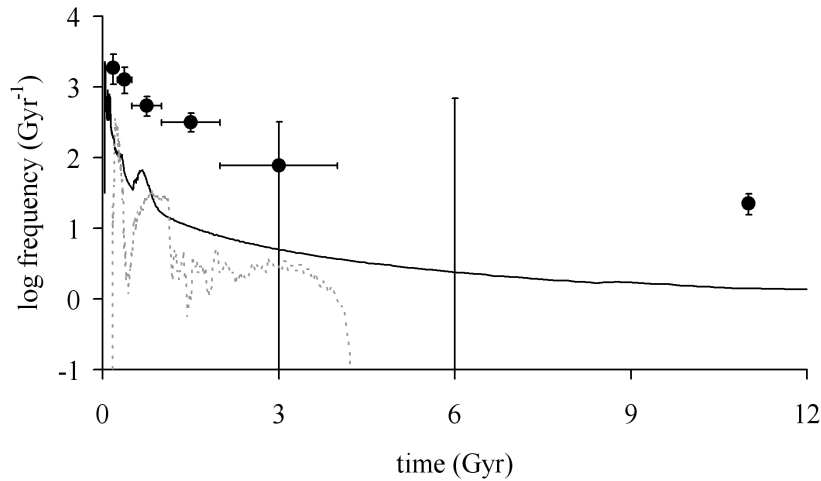}
   \caption{DTDs for $\beta=1$ of the DD (solid black) and SD (dotted gray) scenario using the $\gamma$-formalism for CE. Observational data points of Totani et al. (2008) and Mannucci et al. (2005) (black circles).}
\end{figure*}

\begin{figure*}
   \centering
   \includegraphics[width=10.75cm]{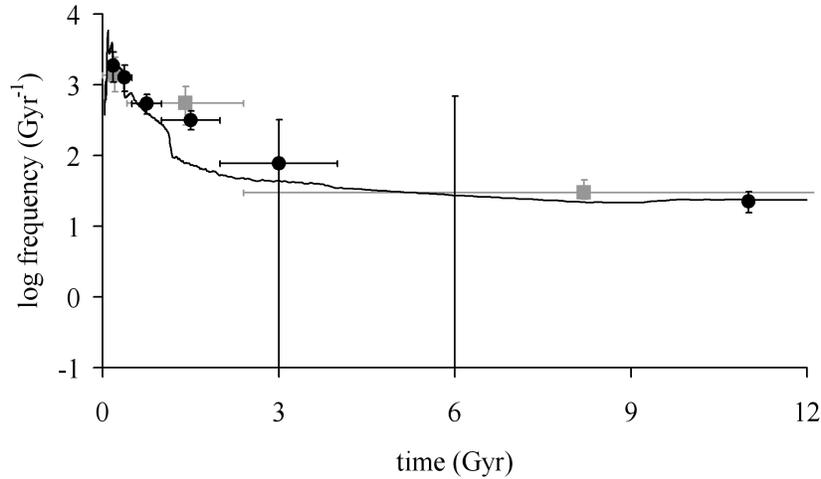}
   \caption{DTD for $\beta=1$ of the SD and DD scenario combined using the $\alpha$-formalism for CE, and with a 10\% increase in MS convective core mass (solid black). Observational data points of Totani et al. (2008) and Mannucci et al. (2005) (black circles), as well as Maoz et al. (2010) (gray squares).}
\end{figure*}

\section{Conclusions}

We find (see also Mennekens et al. 2010) that the single degenerate scenario by itself is incompatible with the morphological shape of the observed delay time distribution of type Ia supernovae. Most double degenerate events are created through a quasi-conservative Roche lobe overflow, followed by a common envelope phase. The resulting critical dependence of the delay time distribution on the mass transfer efficiency during Roche lobe overflow and on the physics of common envelope evolution might be a way to find out more about these processes when more detailed observations become available.

\end{document}